\documentclass[aps,prx,reprint,superscriptaddress,longbibliography,amsmath,amssymb,floatfix]{revtex4-2}

\usepackage{graphicx}
\usepackage[hidelinks]{hyperref}

\graphicspath{{figures/}}

\begin{document}

\title{Machine-accelerated extraction of experimental superconducting parameters for physical insight and prediction across thousands of materials}

\author{Mihir Manium}
\email[Contact author: ]{mihir.manium@physics.ox.ac.uk}
\affiliation{Clarendon Laboratory, Department of Physics, University of Oxford, Oxford, United Kingdom}

\author{Benjamin Geisler}
\affiliation{Department of Physics, University of Florida, Gainesville, Florida, USA}

\author{Peter J. Hirschfeld}
\affiliation{Department of Physics, University of Florida, Gainesville, Florida, USA}

\author{Amalia I. Coldea}
\email[Contact author: ]{amalia.coldea@physics.ox.ac.uk}
\affiliation{Clarendon Laboratory, Department of Physics, University of Oxford, Oxford, United Kingdom}

\begin{abstract}
New superconductors could unlock next-generation technologies from nuclear fusion to lossless electronics. Machine learning can predict candidate superconductors but lacks validated experimental training data beyond the critical temperature. Here we present \textit{AutoSuperCon}, a database of eight superconducting parameters, containing 20{,}354 entries for 6{,}793 superconductors. The data were extracted from over 47{,}000 arXiv preprints using a large language model and validated against reference datasets ($R^2 = 0.96$ with NIMS SuperCon). Using AutoSuperCon, we show that median gap ratios rise from 3.6 in conventional superconductors to 6.4 in cuprates, 28\% of upper critical fields exceed the Pauli limit, and critical temperatures lie below a ceiling proportional to the superfluid density. Further, we train composition-based predictors for the critical temperature ($R^2_{\log}=0.83$) and the upper critical field ($R^2_{\log}=0.69$), boosted by transfer learning. Within known superconducting families, composition alone is sufficient to screen candidates for both parameters. AutoSuperCon thus enables physical insight and data-driven discovery across thousands of superconductors.
\end{abstract}

\maketitle

\section{Introduction}

Discovering new superconductors with desirable properties would advance technologies from high-efficiency electronics to medical imaging to nuclear fusion. However, superconductor discovery has historically owed more to chance than design~\cite{HideoHosono2015}. For example, neither cuprate~\cite{JohannesBednorz1986} nor iron-based~\cite{YoichiKamihara2008} superconductors were predicted by theory. The same is true even for most conventional superconductors, whose phonon-mediated pairing is quantitatively described by the Bardeen--Cooper--Schrieffer (BCS) theory~\cite{JohnBardeen1957} and its extensions. Machine learning offers an alternative route to identifying candidate superconductors \textit{a priori}, with models that can predict the critical temperature $T_{\mathrm{c}}$, the temperature below which a material superconducts, from the chemical composition or crystal structure~\cite{ValentinStanev2018,BRoter2020,TomohikoKonno2021,OmriLesser2025,JasonGibson2025,JasonGibson2026}. Such models are typically trained on materials from the NIMS SuperCon dataset~\cite{NIMS2022}, a compilation of superconductors and their critical temperatures.

\begin{figure*}[t]
    \centering
    \includegraphics{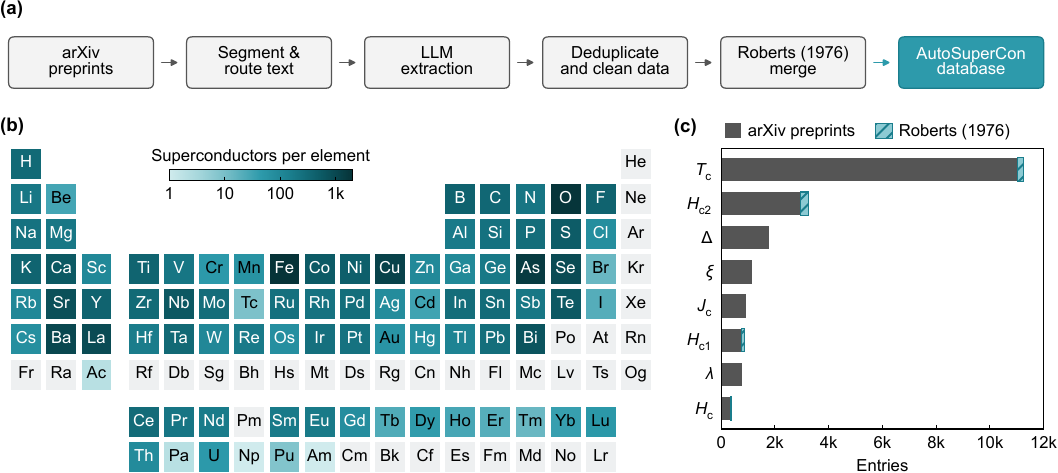}
    \caption{The AutoSuperCon database. (a) Automated workflow from arXiv preprints and the Roberts (1976) survey to the database. (b) Elements of the periodic table coloured by the number of superconductors in the database that contain them. (c) Database entries per superconducting parameter. The Roberts (1976) survey contributes to $T_{\mathrm{c}}$, $H_{\mathrm{c}}$, $H_{\mathrm{c1}}$, and $H_{\mathrm{c2}}$.}
    \label{fig1:ASC}
\end{figure*}

The critical temperature alone, however, is insufficient in practice for identifying useful superconductors. For example, the upper critical field $H_{\mathrm{c2}}$ bounds the fields at which superconducting magnets and cables can operate, and the critical current density $J_{\mathrm{c}}$ restricts the useful current that such conductors can carry~\cite{DavidLarbalestier2001}. First-principles theory can be used to compute these parameters~\cite{BenjaminGeisler2026}, but is so far limited to conventional superconductors and does not extend to unconventional superconductors. Moreover, real materials contain defects and impurities, which are difficult to model, and so these parameters must be measured experimentally.

Experimental superconducting parameters beyond $T_{\mathrm{c}}$ are routinely reported in the literature, but cannot be used to train machine learning models unless tabulated in structured datasets~\cite{AnubhavJain2013,ScottKirklin2015}. While manual curation at this scale involves substantial human labour, the extraction process can be automated~\cite{MatthewSwain2016,JohnDagdelen2024}, and several text-mined superconductor databases have appeared recently~\cite{LucaFoppiano2023,CallumCourt2020,SumanItani2026}. However, these efforts are either restricted to the critical temperature~\cite{LucaFoppiano2023,CallumCourt2020} or largely unvalidated against reference data~\cite{SumanItani2026}.

Here, we develop an automated workflow that uses a large language model (LLM) to extract experimentally measured superconducting parameters from open-access arXiv preprints. Supplemented with our digitisation of the Roberts (1976) survey of pre-arXiv measurements~\cite{BenjaminRoberts1976}, the resulting database, \textit{AutoSuperCon}, comprises 20{,}354 entries for 6{,}793 superconductors, measured at both ambient (75\%) and applied (25\%) pressures, and includes the source(s) of each entry. It contains eight superconducting parameters: the critical temperature $T_{\mathrm{c}}$; the thermodynamic, lower, and upper critical fields $H_{\mathrm{c}}$, $H_{\mathrm{c1}}$, and $H_{\mathrm{c2}}$; the critical current density $J_{\mathrm{c}}$; the coherence length $\xi$; the penetration depth $\lambda$; and the superconducting gap $\Delta$. After validating AutoSuperCon against reference datasets, we first test empirical relations between these parameters and find that gap ratios of unconventional superconductors deviate from weak-coupling BCS theory, upper critical fields frequently violate the Pauli limit, and critical temperatures respect a ceiling proportional to the superfluid density. We then use AutoSuperCon to train composition-based machine learning models to predict critical temperatures ($R^2_{\log}=0.83$) and upper critical fields ($R^2_{\log}=0.69$), boosted by transfer learning from the larger $T_{\mathrm{c}}$ dataset. Neither model extrapolates to superconducting families withheld from training. Within known families, however, composition alone is sufficient to screen candidates for both parameters.

\section{Results and Discussion}

\subsection{Building the AutoSuperCon database}

Our automated workflow (Fig.~\ref{fig1:ASC}(a)) extracts data from the more than 47{,}000 open-access preprints of arXiv's superconductivity category using an LLM. To reduce the number of LLM calls, we split the text of each preprint into segments and pass a segment to the LLM only if it contains relevant keywords, e.g.\ ``critical temperature''. For each superconducting parameter, a system prompt instructs the LLM to return the chemical composition of the material, the measured parameter value, and the experimental conditions, such as the applied pressure, sample form, or magnetic field orientation. Each entry carries the arXiv identifier of its source preprint.

There are multiple possible sources of error in the extracted data, and arXiv preprints are not peer-reviewed. We therefore clean the data by removing implausible values (Methods) and collapse duplicate entries by taking the median parameter value, while recording the number and spread of the contributing reports for each entry. For experimental data from before the arXiv era, we include entries for $T_{\mathrm{c}}$, $H_{\mathrm{c}}$, $H_{\mathrm{c1}}$, and $H_{\mathrm{c2}}$ from the Roberts (1976) survey of superconductor parameters~\cite{BenjaminRoberts1976}. The survey exists as a scanned typewritten report whose tables cannot be read by standard optical character recognition (OCR) software. We therefore transcribe them verbatim with a multimodal LLM and clean the output with a rule-based pipeline (Methods). In this way, AutoSuperCon contains 20{,}354 entries for 6{,}793 superconductors, spanning 84 elements (Fig.~\ref{fig1:ASC}(b)) and eight superconducting parameters~(Fig.~\ref{fig1:ASC}(c)).

\subsection{Validation against reference data}
\begin{figure*}[t]
    \centering
    \includegraphics{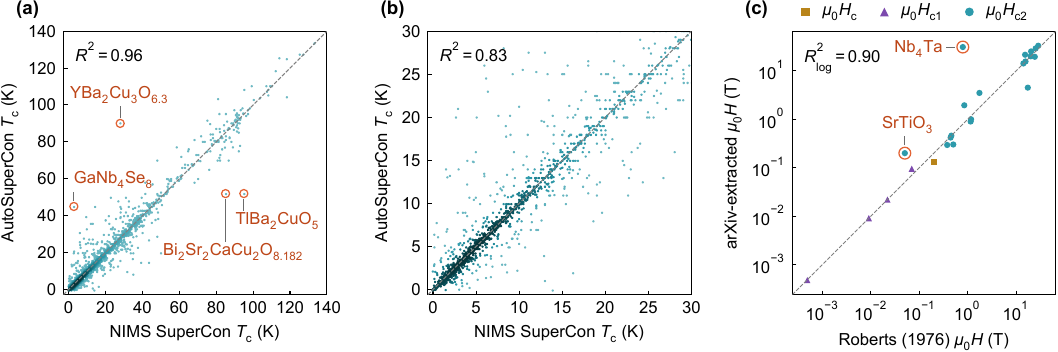}
    \caption{Validation against reference data. (a) AutoSuperCon $T_{\mathrm{c}}$ compared with NIMS SuperCon for the superconductors in both databases ($n = 2{,}439$, $R^2 = 0.96$). (b) The same comparison restricted to NIMS SuperCon $T_{\mathrm{c}} \le 30$~K ($n = 2{,}014$, $R^2 = 0.83$), including 36 points whose AutoSuperCon values lie beyond the displayed range. (c) arXiv-extracted critical fields compared with Roberts (1976) for the 24 values reported by both at the same measurement temperature, within 0.05~K ($R^2_{\log} = 0.90$). In (a) and (b) points are coloured by local point density. Circled outliers in (a) and (c) are labelled by composition.}
    \label{fig2:validation}
\end{figure*}

To quantify the error introduced by our automated extraction workflow, we benchmark the database against reference data (Fig.~\ref{fig2:validation}). For the 2{,}439 superconductors common to AutoSuperCon and NIMS SuperCon, the two databases show close agreement on $T_{\mathrm{c}}$, with $R^2 = 0.96$ (Fig.~\ref{fig2:validation}(a)). The full critical-temperature range is dominated by the cuprates, but the agreement holds even for the 2{,}014 superconductors with $T_{\mathrm{c}} \le 30$~K ($R^2 = 0.83$, Fig.~\ref{fig2:validation}(b)). The worst outliers (circled in Fig.~\ref{fig2:validation}(a)) each trace back to an identifiable failure mode, such as a transient photoinduced value read as an equilibrium $T_{\mathrm{c}}$ or a theoretical value read as a measurement. Such failures are rare, and the median deviation between the two databases is 0.25~K in absolute terms and 3.5\% in relative terms. Disagreements do not always originate with our extraction, since NIMS SuperCon itself is known to contain erroneous records~\cite{KamHamidieh2018,ValentinStanev2018}. We apply the same cleaning rules to every entry and keep these outliers, since selective removal would bias the database.

For the critical fields, the Roberts (1976) survey serves as both a data source and a benchmark. We compare the two sources, the arXiv-extracted data and Roberts (1976), only where both use the same measurement temperature, within 0.05~K. The 24 critical-field values that qualify agree with $R^2_{\log} = 0.90$ (Fig.~\ref{fig2:validation}(c)), where $R^2_{\log}$ is the coefficient of determination computed on $\log_{10}$-transformed values. The two circled outliers are an extraction error and a genuine physical effect, the carrier-density dependence of superconductivity in SrTiO$_3$~\cite{ClementCollignon2019}.

The remaining parameters have no tabulated reference database. For these we compare extracted values with published ranges for well-studied compounds spanning all major families, and all 19 values fall within those ranges. Thus, on every benchmark, errors are small compared with the range each parameter spans, and any individual value can be checked against its source preprint.

\subsection{Testing empirical relations across superconducting families}
\begin{figure*}[t]
    \centering
    \includegraphics{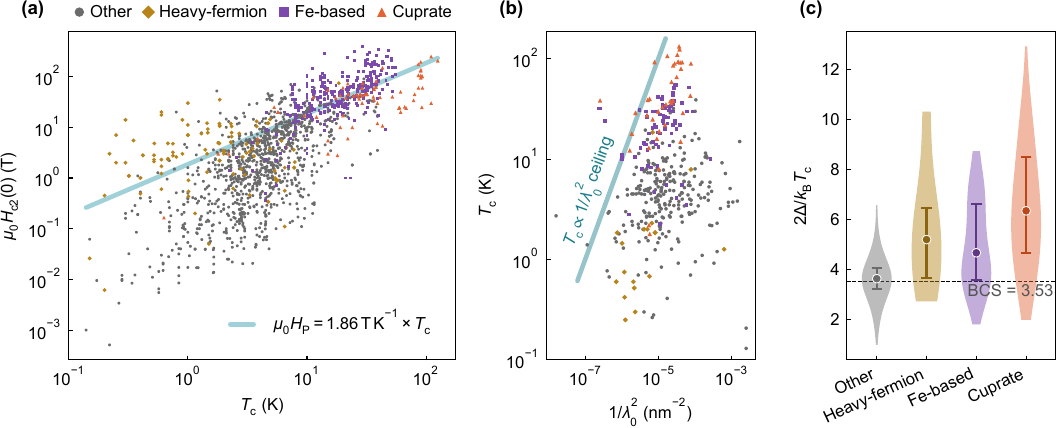}
    \caption{Cross-parameter relations across the superconductor population. (a) Upper critical field as a function of critical temperature for 1{,}554 entries (one per material and magnetic field orientation), of which 28\% exceed the weak-coupling Pauli limit $\mu_0 H_{\mathrm{P}} = 1.86\,\mathrm{T\,K^{-1}} \times T_{\mathrm{c}}$. (b) Critical temperature as a function of the inverse square of the in-plane penetration depth, $1/\lambda_0^2$ (Uemura plot), for 382 superconductors. An empirical ceiling $T_{\mathrm{c}}^{\mathrm{ceil}} \propto 1/\lambda_0^2$ bounds the population, placed for display at the 97.5th percentile of $T_{\mathrm{c}}\lambda_0^2$. (c) Distributions of gap ratios $2\Delta/k_{\mathrm{B}}T_{\mathrm{c}}$ compared with the weak-coupling BCS $s$-wave value of 3.53. Circles mark family medians and whiskers the interquartile range, computed on the full distributions. Violin bodies are truncated at each family's 95th percentile for display.}
    \label{fig3:structure}
\end{figure*}
With multiple experimental parameters for the same materials, AutoSuperCon enables cross-parameter empirical relations to be tested at scale. Figure~\ref{fig3:structure} shows three such relations, each measured over hundreds to thousands of materials. We study bulk samples at ambient pressure, taking critical fields, penetration depths, and gaps in the zero-temperature limit.

Figure~\ref{fig3:structure}(a) shows the upper critical field as a function of the critical temperature for 1{,}554 entries, one per material and magnetic field orientation, spanning four orders of magnitude in field. In weak-coupling BCS theory, upper critical fields are expected to lie below the Pauli limit ($\mu_0 H_{\mathrm{P}} = 1.86\,\mathrm{T\,K^{-1}} \times T_{\mathrm{c}}$)~\cite{AlbertClogston1962,BellurChandrasekhar1962}, above which the energy gained by aligning electron spins with the magnetic field should break singlet Cooper pairs. We find that 28\% of entries exceed it. Violations are more common in the iron-based (66\%) and heavy-fermion (59\%) superconductors than in the cuprates (24\%) and the predominantly conventional ``other'' superconductors (13\%). Such violations are understood to arise from strong coupling~\cite{TerryOrlando1979}, multiband effects~\cite{AlexanderGurevich2011}, spin--orbit scattering~\cite{RichardWerthamer1966}, and non-singlet pairing~\cite{ShengRan2019}.

In the underdoped cuprates, the critical temperature was found to scale with the superfluid density, $T_{\mathrm{c}} \propto 1/\lambda_0^2$, where $\lambda_0$ is the penetration depth at zero temperature~\cite{YasutomoUemura1989}. This Uemura relation was later extended to other exotic superconductors~\cite{YasutomoUemura1991}. Figure~\ref{fig3:structure}(b) shows the Uemura plot for the 382 superconductors in AutoSuperCon with measured in-plane penetration depths, and we find that the relation does not extend to the wider population. Instead, critical temperatures respect an empirical ceiling, $T_{\mathrm{c}}^{\mathrm{ceil}} \propto 1/\lambda_0^2$. Such a ceiling has been rigorously derived for two-dimensional materials, independent of the pairing mechanism~\cite{TamaghnaHazra2019,MohitRanderia2024}, and holds approximately for layered materials with similar interlayer spacings~\cite{ErikaCarlson1999}.

Figure~\ref{fig3:structure}(c) shows distributions of the gap ratio $2\Delta/k_{\mathrm{B}}T_{\mathrm{c}}$, resolved by superconducting family. Each material enters once as the median of its reported gaps, which averages over anisotropy and multiband effects. Weak-coupling BCS theory predicts a ratio of 3.53 for an isotropic $s$-wave gap~\cite{JohnBardeen1957}. As expected, the ``other'' superconductors lie close to this value (median 3.64), while the heavy-fermion (5.20), iron-based (4.68), and cuprate (6.35) distributions all lie above, with their medians exceeding even the ratio of the strongly coupled elemental superconductor Pb ($2\Delta/k_{\mathrm{B}}T_{\mathrm{c}} = 4.5$)~\cite{JulesCarbotte1990}.

Together, these relations provide quantitative metrics against which any future theory of unconventional superconductivity should be compared.

\subsection{Predicting superconducting parameters from chemical composition}
\begin{figure*}[t]
    \centering
    \includegraphics{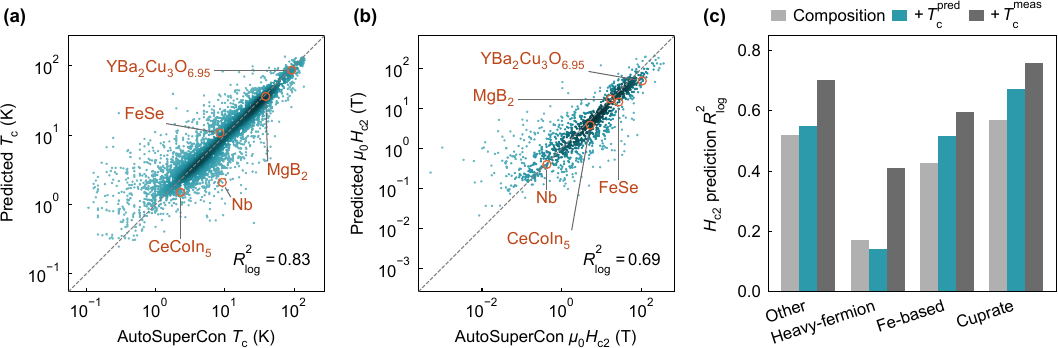}
    \caption{Composition predicts $T_{\mathrm{c}}$ and $H_{\mathrm{c2}}$ for held-out materials. (a) Predicted $T_{\mathrm{c}}$ compared with AutoSuperCon values ($n = 5{,}843$, $R^2_{\log} = 0.83$). (b) Predicted $H_{\mathrm{c2}}$, from composition and the predicted $T_{\mathrm{c}}$, compared with AutoSuperCon values ($n = 1{,}204$, $R^2_{\log} = 0.69$). (c) Per-family $H_{\mathrm{c2}}$ accuracy ($R^2_{\log}$) for models trained on composition only, on composition and the predicted $T_{\mathrm{c}}$, and on composition and the measured $T_{\mathrm{c}}$. All predictions are from TabPFN-3 under five-fold cross-validation. In (a) and (b) points are coloured by local point density, and the same five materials are circled and labelled by composition in both panels.}
    \label{fig4:predictor}
\end{figure*}
To test how well composition alone predicts superconducting parameters, we train models for the critical temperature and the upper critical field (Fig.~\ref{fig4:predictor}). Compositions are represented by Magpie descriptors~\cite{LoganWard2016}, and the model used is TabPFN-3, a tabular foundation model~\cite{NoahHollmann2025,LeoGrinsztajn2026}. Every evaluation metric we report is computed on held-out materials, that is, materials not in the training set.

First, we train a model for critical-temperature prediction on the 5{,}843 superconductors from AutoSuperCon with bulk, ambient-pressure $T_{\mathrm{c}}$ values. The model predicts held-out critical temperatures with $R^2_{\log} = 0.83$ and a median absolute error of 1.6~K (Fig.~\ref{fig4:predictor}(a)). For comparison, the canonical model trained on NIMS SuperCon~\cite{ValentinStanev2018} reaches $R^2_{\log} \approx 0.88$ from a training set of comparable size, but it was trained and evaluated only on materials with $T_{\mathrm{c}} > 10$~K.

Predicting the upper critical field is the harder and more valuable task. It helps determine whether a superconductor is useful for practical applications such as superconducting magnets and cables. To our knowledge, no machine-learned $H_{\mathrm{c2}}$ predictor trained on experimental data has been reported. We therefore train a model on the 1{,}204 superconductors from AutoSuperCon with an orientation-averaged $H_{\mathrm{c2}}$ value and find that predictions from composition alone match the measured values with $R^2_{\log} = 0.67$. Routing the prediction through the critical temperature raises this to $0.69$ (Fig.~\ref{fig4:predictor}(b)). In this two-stage approach, the first stage predicts $T_{\mathrm{c}}$ from composition, and the second predicts $H_{\mathrm{c2}}$ using both composition and this predicted $T_{\mathrm{c}}$ as input. Replacing the predicted $T_{\mathrm{c}}$ with the measured value gives $R^2_{\log} = 0.79$, but only the two-stage approach can be applied to new materials whose critical temperatures have not yet been measured.

The two-stage construction is a form of transfer learning~\cite{SinnoPan2010}, in which information from a data-rich task improves a related data-poor one. Here the $T_{\mathrm{c}}$ dataset is almost five times the size of the $H_{\mathrm{c2}}$ one, and the predicted $T_{\mathrm{c}}$ carries useful information into the $H_{\mathrm{c2}}$ prediction. Because the first stage is trained only on materials outside the $H_{\mathrm{c2}}$ set (Methods), the gain reflects genuine transfer rather than information leakage. The gain in $H_{\mathrm{c2}}$ accuracy is small but statistically resolved ($\Delta R^2_{\log} = +0.022$, 95\% confidence interval $[+0.008, +0.036]$). Figure~\ref{fig4:predictor}(c) shows the per-family accuracies.

\subsection{Limits of composition-based prediction}
\begin{figure}[t]
    \centering
    \includegraphics{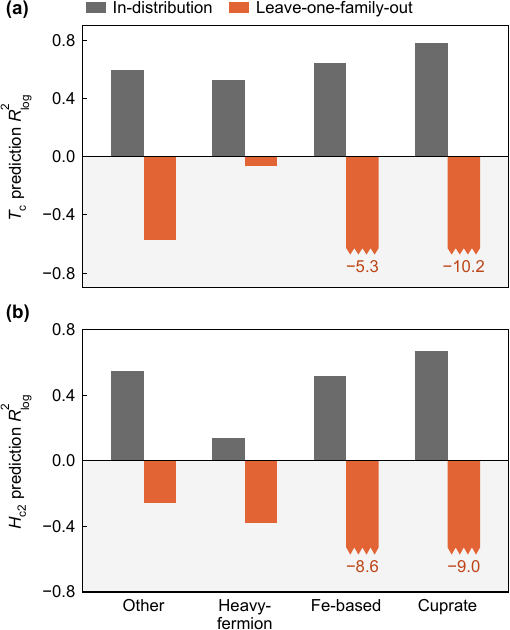}
    \caption{Neither predictor extrapolates to an unseen family. Per-family $R^2_{\log}$ of the TabPFN-3 predictors under random five-fold cross-validation with every family represented in training (in-distribution), and under training with the evaluated family entirely withheld (leave-one-family-out). (a) The $T_{\mathrm{c}}$ predictor. (b) The $H_{\mathrm{c2}}$ predictor, with the family withheld from both the $H_{\mathrm{c2}}$ model and the $T_{\mathrm{c}}$ model that feeds it. Negative $R^2_{\log}$ means that the predictions are worse than guessing the family mean. Off-scale bars are truncated and labelled with their values.}
    \label{fig5:generalisation}
\end{figure}
In our machine learning analyses so far, predictions of $T_{\mathrm{c}}$ and $H_{\mathrm{c2}}$ are made with every superconducting family represented in the training data. The historic breakthroughs, however, were discoveries of wholly new families, and so the harder test is whether a model can predict an unseen class of superconductors~\cite{BryceMeredig2018}. We therefore perform a leave-one-family-out evaluation, withholding an entire family from training and predicting it. Figure~\ref{fig5:generalisation} shows that both the $T_{\mathrm{c}}$ and $H_{\mathrm{c2}}$ predictors fail, most severely for the cuprate and iron-based families.

When withheld from training, the cuprates are predicted at $R^2_{\log} = -10.2$ for $T_{\mathrm{c}}$ (Fig.~\ref{fig5:generalisation}(a)), compared with $0.78$ when present in training, and the iron-based family behaves similarly. The models are extrapolating conventional temperature and magnetic field scales onto materials that do not obey them. Predictions for heavy-fermion and ``other'' superconductors degrade more gently. Cuprates and iron-based materials form tight clusters in composition space, far from every other superconductor, so withholding one leaves no nearby material to learn from. Meanwhile, heavy-fermion and ``other'' materials lie closer to the rest of the population.

These findings expose a limitation of composition-based learning, one that could be partially overcome with more experimental training data. For example, when we treat the cuprates as a newly discovered family, with their $H_{\mathrm{c2}}$ data withheld but their critical temperatures known, the $H_{\mathrm{c2}}$ prediction improves from $R^2_{\log} = -9.0$ (Fig.~\ref{fig5:generalisation}(b)) to $+0.4$. Within known regions of composition space, where most materials optimisation happens, the predictors are accurate screening tools.

\section{Conclusion}

We constructed AutoSuperCon, a database of experimental superconducting parameters, using a large language model to extract data from arXiv preprints and historical literature. The database contains 20{,}354 entries covering 6{,}793 superconductors. It enables large-scale analysis of superconducting parameters and reveals quantitative deviations from weak-coupling BCS theory, including enhanced superconducting gap ratios and upper critical fields exceeding the Pauli limit. We also identified an empirical upper bound on critical temperatures linked to the superfluid density. Finally, the database allows the training of machine learning models that predict critical temperatures and upper critical fields directly from chemical composition, demonstrating the potential of AI-driven approaches to accelerate superconducting materials research.

Our automated extraction workflow is not specific to superconductivity and can be applied to frustrated magnets, topological matter, or any other class of quantum materials to build structured experimental databases. The ultimate goal is to have models trained on high-quality, reproducible data to predict experimental parameters and thereby guide the search for novel functional materials.

\section{Methods}
\label{sec:methods}

\subsection{Corpus and LLM extraction}
We downloaded all preprints from the arXiv \texttt{cond-mat.supr-con} category posted up to March 2026 and parsed the text with PyMuPDF (v1.26.7)~\cite{Artifex2025}. Each preprint's text was split into overlapping segments of 2{,}500 characters with a 500-character overlap, breaking preferentially at paragraph and sentence boundaries. We only considered segments if they contained a parameter keyword, a number with a compatible unit, and the substring ``superconduct''. These segments were then processed by an LLM (OpenAI \texttt{gpt-5-mini}~\cite{OpenAI2025}, 2025-08-07 snapshot) in strict structured-output mode. For each parameter, a system prompt with worked examples and a JSON schema instructed the model to return zero or more reported parameter values per segment. Each entry contains the chemical formula of the superconductor, the parameter value in a unit fixed by the schema, the experimental method, the sample form (e.g.\ thin film or bulk), and parameter-specific conditions such as the field orientation, measurement temperature, or applied pressure.

\subsection{Cleaning and deduplication}
We parsed chemical formulae with pymatgen~\cite{ShyuePingOng2013} and standardised these with alphabetised elements and normalised coefficients. Formulae with unresolved variables (e.g.\ FeSe$_{1-x}$S$_{x}$) were rejected. Implausible values were removed based on per-parameter allowed ranges. Further, entries for formulae with a noble gas were removed. For duplicate reports of the same formula under the same measurement conditions, we took the median value. Reports at an unspecified non-ambient pressure were pooled only within a single preprint. Cross-parameter consistency filters removed entries violating inequalities and ratio bounds (e.g.\ $H_{\mathrm{c1}} < H_{\mathrm{c}} < H_{\mathrm{c2}}$ for type-II superconductors).

\subsection{Digitisation of the Roberts (1976) survey}
Table 5 of the Roberts (1976) survey~\cite{BenjaminRoberts1976} is available only as a scan of a typewritten report that standard OCR cannot read. We passed images of the scanned pages to a multimodal LLM (Google Gemini 3 Flash)~\cite{GoogleDeepMind2025}, instructed only to convert each image into a CSV table. The returned tables reproduce the printed columns and preserve units, sample-form notes, and anisotropy markers as printed. We used a deterministic pipeline to clean the 605 transcribed rows, converting units (1~kOe $= 0.1$~T), standardising formulae, recording the observation temperature, and collapsing duplicates of the same formula, temperature, and orientation to their median. This yielded 404 material rows, of which 363 are bulk and 41 are thin films. The thin-film rows are labelled as such and excluded from subsequent analyses, as with the rest of AutoSuperCon. We only add Roberts (1976) rows where they do not duplicate an equivalent arXiv-extracted entry, totalling 639 entries (197 for $T_{\mathrm{c}}$, 46 for $H_{\mathrm{c}}$, 87 for $H_{\mathrm{c1}}$, and 309 for $H_{\mathrm{c2}}$), with 254 new superconductors not present in the arXiv-extracted data.

\subsection{Validation}
For comparing databases, we match entries based on their standardised formulae. Against NIMS SuperCon we compare per-formula critical temperatures and apply the same cleaning procedures described above to both datasets. For critical-field values we compare arXiv-extracted and Roberts (1976) data before merging, pairing values only where the two sources report the same measurement temperature to within 0.05~K. We further check parameter values for well-studied compounds against published ranges.

\subsection{Data selection and families}
For all quantitative analyses, we use bulk samples at ambient pressure, with critical fields, gaps, and penetration depths taken at zero temperature. The Uemura analysis uses in-plane ($ab$) penetration depths, with direction-unspecified entries where no $ab$ measurement exists. Gap ratios use the directly reported ratio where given and otherwise the calculated ratio $2\Delta/k_{\mathrm{B}}T_{\mathrm{c}}$. We assign materials to families using their compositions, with the following rules applied in order. Iron-based requires Fe plus a pnictogen or chalcogen from \{As, P, S, Se, Te\}. Cuprate requires Cu and O. Nickelate requires Ni and O, excluding Ni-pnictide and Ni-chalcogenide compounds. Hydride requires H, D, or T. Heavy-fermion requires one of Ce, Pr, Yb, U, Np, or Pu, the elements whose partially filled, hybridising $f$ shells produce heavy-fermion behaviour~\cite{ChristianPfleiderer2009}. Everything else is ``other''. Hydrides are too few to resolve separately for ambient-pressure analyses and are shown with ``other''.

\subsection{Machine learning models}
As input to machine learning models, we represent compositions numerically using 132 Magpie descriptors~\cite{LoganWard2016} computed with matminer~\cite{LoganWard2018}. For the machine learning analyses, we use the TabPFN-3~\cite{NoahHollmann2025,LeoGrinsztajn2026} model. We also use a 300-tree random forest~\cite{LeoBreiman2001} as a cross-check, implemented in scikit-learn~\cite{FabianPedregosa2011}. Targets are $\log_{10}$-transformed and $R^2_{\log}$ is computed relative to the $y=x$ line in log space. For all reported scores, we use held-out folds of a five-fold cross-validation, with confidence intervals from 10{,}000 bootstrap resamples of the held-out materials. We assess the gains from transfer learning with paired differences, in which both descriptor sets score the same materials and the difference in $R^2_{\log}$ is bootstrapped over jointly resampled materials (10{,}000 resamples). For machine learning analysis of $H_{\mathrm{c2}}$, we average over orientations using a measurement-weighted median per material, so no composition appears on both sides of a fold split. Entries without a measured $T_{\mathrm{c}}$ are dropped, leaving 1{,}204 materials. The first stage of the two-stage model is a single $T_{\mathrm{c}}$ fit on the 4{,}639 materials of the $T_{\mathrm{c}}$ cohort that are not among these 1{,}204, and so its predicted $T_{\mathrm{c}}$ is out of sample for every $H_{\mathrm{c2}}$ material. It is a separate $T_{\mathrm{c}}$ predictor from that of Fig.~\ref{fig4:predictor}(a).

\begin{acknowledgments}
We thank Steven Simon for scientific advice. M.M. acknowledges support from the EPSRC Centre for Doctoral Training in Superconductivity [EP/Y035453/1], the Clarendon Fund Scholarship, and the New College Graduate Award in MPLS at the University of Oxford. A.I.C. acknowledges support from an EPSRC Open Fellowship [UKRI3274], the EPSRC Centre for Doctoral Training in Superconductivity [EP/Y035453/1], and the Oxford Centre for Applied Superconductivity. B.G. and P.J.H. acknowledge funding from Grant No.~NSF-DMREF-2522891. P.J.H. is grateful for the partial support of the Leverhulme Trust during the research for this paper.
\end{acknowledgments}

\bigskip

M.M. and A.I.C. conceived the study. M.M. built the extraction pipeline, assembled the database, and performed the analysis and the machine learning predictions. M.M. and A.I.C. wrote the paper. B.G. and P.J.H. contributed the Roberts (1976) source material and the Pauli-limit analysis. A.I.C. supervised the research and secured funding. All authors discussed the results and approved the manuscript.

\bibliography{references}

\end{document}